%% file: main.tex
\pdfoutput=1
\documentclass[aps,prl,reprint,amsmath,amssymb]{revtex4-2}

\usepackage[T1]{fontenc}
\usepackage{lmodern}
\usepackage{graphicx}
\usepackage{bm}
\usepackage{xcolor}
\usepackage[colorlinks=true,linkcolor={blue!60!black},citecolor={blue!60!black},urlcolor={blue!60!black}]{hyperref}

\newcommand{\tr}{\operatorname{tr}}
\newcommand{\RR}{\mathbb{R}}
\newcommand{\ZZ}{\mathbb{Z}}
\newcommand{\lb}{\bar\lambda}
\newcommand{\Rh}{\widehat R}
\newcommand{\htl}{\tilde h}
\newcommand{\ftl}{\tilde f}
\newcommand{\ztl}{\tilde \zeta}
\newcommand{\nf}[1]{N_{f,#1}}
\newcommand{\cT}{\mathcal{T}}

\newcommand{\figfile}{figs/fig1.pdf}
\begin{document}

\title{A Superfluid at All Temperatures}

\author{Sven Harder}
\author{Adar Sharon}
\affiliation{Mani L. Bhaumik Institute for Theoretical Physics, Department of Physics and Astronomy,\\
University of California, Los Angeles, CA 90095, USA}

\begin{abstract}
We construct unitary, local, 3+1-dimensional conformal field theories with finitely many fields whose thermal state breaks a continuous global symmetry at every temperature. They are finite-$N$ versions of the Chaudhuri--Choi--Rabinovici theories, whose planar circle of fixed points has an ordered arc. The circle is lifted at order $1/N^2$, and an infinite family of fixed points on this arc reaches arbitrarily weak coupling, where the superfluid is under parametric control. We also show that weakly gauging the symmetry gives a superconductor over an exponential range of temperatures.
\end{abstract}

\maketitle

\emph{Introduction.}---Ordered phases of matter usually melt when they are heated. The equilibrium state minimizes the free energy $E-TS$, and at high temperature the entropy of disordered configurations dominates. In relativistic field theory this expectation is known as symmetry restoration~\cite{KirzhnitsLinde:1972,DolanJackiw:1974,Weinberg:1974}. Weinberg pointed out that it is not a theorem~\cite{Weinberg:1974}; with several scalar multiplets, a negative cross-coupling can make one of the thermal masses negative while the potential remains bounded from below, such that a symmetry stays spontaneously broken at arbitrarily high temperature. This possibility has been explored extensively, for instance in cosmology~\cite{MohapatraSenjanovic:1979b,LangackerPi:1980,DvaliMelfoSenjanovic:1995,MeadeRamani:2019}, but two major obstacles have accompanied it. The scalar theories originally considered are not ultraviolet complete: their couplings grow at short distances, so they make sense only up to a cutoff scale, and hence a maximal temperature~\cite{Orloff:1997}. Furthermore, the effect results from a competition between loop contributions of opposite sign, which higher-order corrections might undo~\cite{BimonteLozano:1995a,Roos:1996,PintoRamos:2000}.

Conformal field theories (CFTs) remove the first obstacle. On $\RR^3\times S^1_\beta$ the temperature $T=1/\beta$ is the only scale. If the thermal state of a CFT breaks a symmetry at one temperature, it breaks it at all temperatures. This observation motivated the study of thermal order in CFTs~\cite{Chai:2020a,Chai:2020b}, which is now understood as an instance of entropic order~\cite{Han:2026}: ordering one set of degrees of freedom allows another set to fluctuate more, so that the ordered state has the larger entropy. As emphasized in~\cite{KomargodskiPopov:2025}, in quantum field theory symmetry breaking at all temperatures had previously been established only in a non-integer number of dimensions, with non-local interactions, or in the strict planar limit. Each of these idealizations gives up a standard property of quantum field theory. Scalar theories in $d=4-\epsilon$ are not unitary~\cite{HogervorstRychkovvanRees:2016}, long-range models~\cite{ChaiDymarskySmolkin:2022,Chai:2022} are not local, and the planar limit has infinitely many fields at each spacetime point. The last idealization is not innocent: conclusions drawn at $N=\infty$ can violate constraints that hold at any finite $N$, because the large-$N$ limit need not commute with the long-distance limit~\cite{Witten:1978,RabinoviciSmolkin:2011}. In the two-dimensional $SU(N)$ Thirring model the $1/N$ expansion appears to predict spontaneous breaking of the continuous chiral symmetry, which is impossible~\cite{Coleman:1973}; instead, the correlator of the order parameter decays as $|x|^{-1/N}$, which becomes long-range order only at $N=\infty$~\cite{Witten:1978}. All idealizations were first avoided in~\cite{KomargodskiPopov:2025}, where a unitary, local CFT with finitely many fields breaks its discrete $\ZZ_2$ symmetry spontaneously at all temperatures in $2+1$ dimensions, with control at large but finite $N$; see also~\cite{HawashinRongScherer:2025,ChaiRabinoviciSinhaSmolkin:2021,SmolkinYung:2026a,SmolkinYung:2026b,ChaudhuriHawashinRabinoviciScherer:2026}.

Breaking a continuous symmetry at nonzero temperature requires going to $3+1$ dimensions, where the thermal state lives on $\RR^3$. Chaudhuri, Choi, and Rabinovici (CCR) constructed weakly coupled conformal gauge theories whose planar limit has a circle of fixed points, an arc of which breaks a $U(1)$ symmetry at all temperatures~\cite{CCR:2021,ChaudhuriRabinovici:2021}. Whether an ordered fixed point survives at finite $N$ was left open~\cite{NoteAF}.

In this Letter we show that ordered fixed points do survive at finite $N$. The circle is lifted at order $1/N^2$ and leaves behind a discrete set of fixed points, whose positions on the circle depend on the ranks of the gauge groups and the numbers of flavors. These can be chosen so that a fixed point lies on the ordered arc. We construct an infinite family of such theories, in which the ranks grow and all couplings become arbitrarily weak. In this family the existence of an ordered fixed point, and the fact that its thermal state is a relativistic superfluid at every temperature, are under parametric perturbative control. We also show that coupling the broken symmetry to a weakly coupled photon gives a superconductor over an exponentially large range of temperatures~\cite{NoteKP}.

\begin{figure*}[t]
\centering
\includegraphics[width=\textwidth]{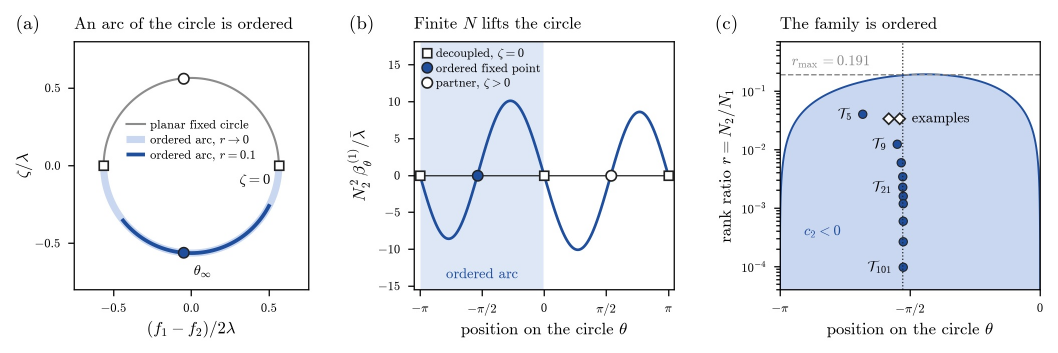}
\caption{From the planar conformal manifold to ordered CFTs at finite $N$.
(a) The circle of planar fixed points~(\ref{eq:circle}) in the plane of the double-trace couplings $(f_1-f_2)/2$ and $\zeta$, in units of $\lambda$.
Its lower half, where $\zeta<0$, is ordered ($c_2<0$) for $r=N_2/N_1\to0$.
At $r=0.1$ the ordered arc is shorter, and it shrinks to a point at $r=r_{\max}=0.191$.
The markers show the points that survive at finite $N$ in the family $\cT_n$ as $n\to\infty$: the decoupled points with $\zeta=0$ (squares), the ordered fixed point at $\theta_\infty$ (filled circle), and its partner at $-\theta_\infty$, which has $\zeta>0$ and is not ordered (open circle).
(b) The $\beta$-function $\beta^{(1)}_\theta$~(\ref{eq:L}) for the motion along the old circle at finite $N$, for $\cT_{21}$.
Its zeros are the fixed points that survive.
The shaded band is the ordered arc for $r=1/441$.
(c) The ordered region $c_2<0$ in the plane of $\theta$ and $r$ (shaded), with the two-loop fixed points of the family $\cT_n$ for $5\le n\le101$, at $r=1/n^2$ (circles), and of the theories $(N_1,\nf1,N_2,\nf2)=(150,783,5,26)$ and $(300,1562,10,52)$, at $r=1/30$ (diamonds, right and left).
The dotted line marks $\theta_\infty$.}
\label{fig:main}
\end{figure*}

\emph{The theories.}---The gauge group is $G_1\times G_2$, with $G_i=SU(N_i)_{\rm a}\times SU(N_i)_{\rm b}$ and $N_2<N_1$. The sectors $i=1,2$ each contain $\nf i$ Dirac fermions $\psi_i$ in the fundamental of $SU(N_i)_{\rm a}$, $\nf i$ Dirac fermions $\chi_i$ in the fundamental of $SU(N_i)_{\rm b}$, and a complex scalar $\Phi_i$ in the bifundamental. A $\ZZ_2$ symmetry exchanges the two factors and equates their gauge couplings to a common value $g_i$, and all Yukawa couplings are forbidden by flavor symmetries. The two sectors interact only through the scalar potential
\begin{equation}\label{eq:V}
\begin{split}
V={}&\sum_{i=1,2}\Big[\htl_i\tr\big(\Phi_i^\dagger\Phi_i\big)^2+\ftl_i\big(\tr\Phi_i^\dagger\Phi_i\big)^2\Big]\\
&+2\ztl\,\tr\Phi_1^\dagger\Phi_1\,\tr\Phi_2^\dagger\Phi_2\,.
\end{split}
\end{equation}
We use the 't~Hooft couplings $\lambda_i=g_i^2N_i/16\pi^2$, $h_i=N_i\htl_i/16\pi^2$, $f_i=N_i^2\ftl_i/16\pi^2$, and $\zeta=N_1N_2\ztl/16\pi^2$.

The symmetry of interest is $U(1)_{\Phi_2}\colon\Phi_2\to e^{i\alpha}\Phi_2$. It is anomaly-free and acts faithfully on gauge-invariant operators, since $\bar\psi_2\Phi_2\chi_2$ has charge one. Its gauge-invariant order parameter $\det\Phi_2$ has charge $N_2$.

Each sector has a Banks--Zaks fixed point~\cite{Caswell:1974,BanksZaks:1982} when $x_i\equiv\nf i/N_i$ is slightly below $21/4$. With $m_i\equiv21N_i-4\nf i$, the two-loop value is
\begin{equation}\label{eq:BZ}
\lambda_i^*=\frac{21-4x_i}{26x_i-54-3(4+2x_i)/N_i^2}\simeq\frac{m_i}{82.5\,N_i}\,.
\end{equation}
The quartic couplings first enter the $\beta$-functions of the gauge couplings at three loops, such that we can treat $\lambda_1^*$ and $\lambda_2^*$ as fixed in the $\beta$-functions of the quartic couplings.

\emph{The planar circle.}---In the planar limit the single-trace couplings sit at $h_i=(3-\sqrt6)\lambda_i/4$. When $\lambda_1=\lambda_2=\lambda$, the double-trace couplings can be assembled into a symmetric matrix $F$ with diagonal entries $f_1,f_2$ and off-diagonal entry $\zeta$, obeying~\cite{CCR:2021}
\begin{equation}\label{eq:betaF}
\beta_F=\big(12h^2+\tfrac92\lambda^2\big)\mathbf 1+(16h-12\lambda)F+4F^2\,.
\end{equation}
This is covariant under $F\to OFO^T$ with $O\in O(2)$, which is an accidental symmetry of planar double-trace dynamics that rotates $\tr\Phi_1^\dagger\Phi_1$ into $\tr\Phi_2^\dagger\Phi_2$. The fixed points form an $S^1$ conformal manifold
\begin{equation}\label{eq:circle}
\tfrac12(f_1+f_2)=\tfrac{\sqrt6}{2}\,\lambda\,,\qquad \tfrac12(f_1-f_2)+i\zeta=\Rh\lambda\,e^{i\theta}\,,
\end{equation}
with $\Rh=[(18\sqrt6-39)/16]^{1/2}\approx0.564$ [Fig.~\ref{fig:main}(a)]. The circle persists to all orders in planar perturbation theory~\cite{CCR:2021}, as follows from the fact that planar double-trace $\beta$-functions are quadratic in the double-trace couplings, with coefficients fixed by the single-trace data~\cite{DymarskyKlebanovRoiban:2005,PomoniRastelli:2009}.

Whether the thermal state is ordered is determined by the effective potential of the static scalar modes on $\RR^3\times S^1_\beta$. At leading order it is $V+16\pi^2T^2\sum_ic_i\tr\Phi_i^\dagger\Phi_i$, with one-loop thermal masses $c_i$ linear in the couplings. Since $T$ is the only scale, a negative $c_i$, with $V$ being bounded from below, destabilizes the symmetric point at every temperature. For $h_i>0$, $V$ is bounded if and only if $a_i\equiv h_i+f_i>0$ and $\zeta>-\sqrt{a_1a_2}$, which holds on the entire $S^1$. On the circle
\begin{equation}\label{eq:c2}
\frac{c_2}{\lambda}=\frac12+\frac{\Rh}{6}\Big(\frac{\sin\theta}{r}-\cos\theta\Big)\,,\qquad r\equiv\frac{N_2}{N_1}\,,
\end{equation}
while $c_1/\lambda=\frac12+\frac{\Rh}{6}(\cos\theta+r\sin\theta)>0$. The $\beta$-functions do not depend on $r$, but $c_2$ does: each of the $N_1^2$ complex components of $\Phi_1$ contributes $T^2/12$ to $\langle\tr\Phi_1^\dagger\Phi_1\rangle$, and the double-trace coupling turns this into a mass for $\Phi_2$ enhanced by $N_1/N_2$. Thus $c_2<0$ on the arc $-\sin\theta>r(3/\Rh-\cos\theta)$, where $\zeta<0$, which exists for $r<0.19144$~\cite{CCR:2021}. On this ordered arc $\Phi_2$ condenses at every temperature.

\emph{Lifting at order $1/N^2$.}---At finite $N$ the accidental $O(2)$ is broken in two ways: by non-planar terms in the $\beta$-functions, and by the mismatch $\delta=\lambda_1^*-\lambda_2^*$ between the Banks--Zaks couplings, which are rational numbers determined by $(N_i,\nf i)$. There is no correction of order $1/N$. Cutting a diagram at its double-trace vertices decomposes it into single-trace blocks. A block with $k$ insertions of $\tr\Phi_i^\dagger\Phi_i/N_i$ scales as $N_i^{2-k}$, with a further $N_i^{-2}$ for each handle, and since every double-trace vertex supplies two insertions, blocks with odd $k$ occur in pairs. The ranks therefore enter only through $1/N_i^2$ and $1/(N_1N_2)$, as the explicit one- and two-loop $\beta$-functions confirm (End Matter). When $\delta$ is of order $\lb/N_2^2$, with $\lb\equiv(\lambda_1^*+\lambda_2^*)/2$, the conformal manifold is therefore lifted at order $1/N^2$, the first order beyond the planar limit.

Near the circle, the transverse components of the finite-$N$ $\beta$-functions can be set to zero by shifting the couplings slightly off the circle. No analogous shift can eliminate the component tangent to the circle. The surviving fixed points are therefore determined by the zeros of $\beta^{(1)}_\theta$, the leading large-$N$ contribution to the angular $\beta$-function $\beta_\theta=\frac{d\theta}{d\log\mu}$, which governs the RG flow along the circle (End Matter). We find
\begin{equation}\label{eq:L}
\beta^{(1)}_\theta=\frac{\lb}{2\Rh N_2^2}\,\sin\theta\,\big[\Delta+g_r(\theta)\big]\,,\qquad \Delta\equiv\kappa\,\frac{\delta\,N_2^2}{\lb}\,,
\end{equation}
with $\kappa=(18\sqrt6-39)/2\approx2.545$ and $g_r$ given in Eq.~(\ref{eq:gr}). For $r\to0$, $g_0=\kappa\cos^2\theta+b\cos\theta+a$, with $a\approx29.97$ and $b\approx-20.94$; it increases monotonically from $11.58$ at $\theta=0$ to $53.45$ at $\theta=\pm\pi$. The zeros at $\sin\theta=0$ are the decoupled theories with $\zeta=0$. Interacting fixed points exist if and only if
\begin{equation}\label{eq:window}
-53.45<\Delta<-11.58\,.
\end{equation}
In that case, there are two solutions at $\theta=\pm\theta_*$ [Fig.~\ref{fig:main}(b)], where the direction along the old circle is irrelevant, with scaling dimension $4+O(\lb/N_2^2)$. The one with $\zeta<0$ lies on the ordered arc.

\emph{The integers as a tuning knob.}---CCR achieved $\lambda_1^*=\lambda_2^*$ in the planar limit by setting $x_1=x_2$. At finite $N$ no such tuning is available, but none is needed. Eq.~(\ref{eq:BZ}) gives
\begin{equation}\label{eq:Delta-int}
\Delta\simeq\kappa\bigg[\frac{N_2^3}{m_2}\Big(\frac{m_1}{N_1}-\frac{m_2}{N_2}\Big)-\frac{87}{165}\bigg]\,,
\end{equation}
which can be placed in the window~(\ref{eq:window}) by a choice of integers. Since $m_1$ changes in steps of four, this requires $N_1\gtrsim0.24\,N_2^3/m_2$. The granularity of the integers, rather than the bound $r<0.19$, forces the large sector to be much larger than the small one. Two explicit examples are $(N_1,\nf1,N_2,\nf2)=(150,783,5,26)$ and $(300,1562,10,52)$, with $\lambda_i^*$ between $1.4\times10^{-3}$ and $2.6\times10^{-3}$. The fixed points of their one- and two-loop $\beta$-functions, exact in $N_i$, are ordered (Table~\ref{tab:fp}).

\emph{An infinite family.}---The bound $N_1\gtrsim0.24\,N_2^3/m_2$ suggests $N_2=n$ and $N_1=n^3$. Integrality of $\nf1$ and $\nf2$ then leads to the theories $\cT_n$,
\begin{equation}\label{eq:family}
\begin{aligned}
N_1&=n^3\,, & \nf1&=\tfrac14\big(21n^3-n^2+12\big)\,,\\
N_2&=n\,, & \nf2&=\tfrac14\big(21n-1\big)\,,
\end{aligned}
\end{equation}
with $n\equiv1\pmod4$. Along this family, the coupling $\lambda_i^*\simeq2/(165n)$ becomes parametrically small and $\Delta\to\Delta_\infty=-\frac{689}{55}\kappa\approx-31.89$, which sits inside the allowed window. The interacting fixed point approaches $\theta_\infty\approx-1.662$, where $h_i/\lb\to0.138$, $f_1/\lb\to1.174$, $f_2/\lb\to1.276$, and $\zeta/\lb\to-0.562$ [Fig.~\ref{fig:main}(c)]. For ten members with $5\le n\le101$ we have verified that the one- and two-loop $\beta$-functions, exact in $N_i$, have this fixed point, with a potential bounded below, $c_1>0$, and $c_2<0$ (six of them are listed in Table~\ref{tab:fp}).

The family turns these results into statements that hold for all sufficiently large $n$. Along it all couplings are of order $1/n$, $r=1/n^2$, and the terms that break the $O(2)$ are of order $1/n^4$ at one loop; since the circle has radius $\Rh\lb$, they give $\beta^{(1)}_\theta$ of order $1/n^3$. Every further loop adds at least one power of $1/n$, and since the ranks enter only through $1/N_i^2$ and $1/(N_1N_2)$, no inverse powers of $r$ appear. Hence
\begin{equation}
\beta_\theta=\frac{1}{n^3}\Big[\frac{1}{165\Rh}\sin\theta\,\big(\Delta_\infty+g_0(\theta)\big)+O(1/n)\Big].
\end{equation}
Since $g_0$ is monotonic, the leading term has a simple zero at $\theta_\infty$, and a simple zero cannot be removed by a small perturbation. The transverse $\beta$-functions, whose Jacobian has eigenvalues of order $1/n$, can be solved near the circle, and for all sufficiently large $n$ the full $\beta_\theta$ then has a zero near $\theta_\infty$. This is the same argument that establishes the Banks--Zaks fixed point at large $N$.

It is essential that the argument applies to the entire family. For each individual theory the existence of the ordered fixed point rests on numerical evidence, as it does \textit{e.g.}~for the Banks--Zaks fixed point of $SU(3)$ with sixteen flavors. Here, it is more delicate than usual: the position on the circle is set by $\beta^{(1)}_\theta$, which is of order $\lb/N_2^2$, and the two-loop terms shift it by about $50\lb$, a large coefficient. Along the family this shift, approximately $0.6/n$, vanishes, as do all higher-loop corrections relative to the leading term. The ordered fixed points of the family are therefore under parametric perturbative control.

\emph{The thermal state.}---At the ordered fixed point, the leading-order thermal effective potential can be minimized exactly. Since $h_i>0$, at fixed $\tr\Phi_i^\dagger\Phi_i$ the single-trace term is smallest when all eigenvalues of $\Phi_i^\dagger\Phi_i$ are equal. For $c_2<0<c_1$ the global minimum is then
\begin{equation}\label{eq:min}
\Phi_1=0\,,\qquad \Phi_2=v\,U\,,\qquad v^2=-\frac{c_2N_2}{2a_2}\,T^2\,,
\end{equation}
with $U\in U(N_2)$, provided that $c_1N_1a_2>\zeta c_2N_2$, which holds with a wide margin. Relative to the symmetric point the free energy density is $\Delta f_*=-4\pi^2c_2^2N_2^2T^4/a_2$, and along the family we find $\Delta f_*\to-0.0029673\,n^5T^4$.

The condensate breaks $U(1)_{\Phi_2}$ to $\ZZ_{N_2}$. A rotation $e^{i\alpha}$ can be undone by a gauge transformation only if $e^{iN_2\alpha}=1$, since gauge transformations leave $\det\Phi_2$ invariant. Accordingly $\langle\det\Phi_2\rangle_T=v^{N_2}\det U\neq0$, and the phase of $\det\Phi_2$ is a gauge-invariant Goldstone boson of the three-dimensional theory of static modes. The gauge group is Higgsed to $SU(N_2)_{\rm diag}$, as in color--flavor locking~\cite{AlfordRajagopalWilczek:1999}, but here both factors are gauged, and the only broken global symmetry is $U(1)_{\Phi_2}$. The thermal state is a relativistic superfluid~\cite{Son:2002}, with stiffness $2v^2/(N_2T)\simeq0.066\,n^2T$.

\emph{Beyond leading order.}---The ordered minimum results from a competition between the tree-level potential and one-loop thermal masses, which might be undone by higher orders as explained in the introduction. This does not happen here. At the minimum, the masses in the small sector grow like $\sqrt n\,T$ (for instance, the Higgsed vectors have $M_W^2\simeq0.127\,nT^2$), but its contributions to $c_2$ are suppressed by $r$ relative to those of the large sector. The $2n^6$ real scalars of the large sector stay light, with thermal mass $M_{1,0}\simeq0.97\,T/\sqrt n$ at $\Phi_2=0$, so the leading correction to $\Delta f_*$ comes from their static modes. Their free energy $-2N_1^2TM_1^3/(12\pi)$ depends on the condensate through $M_1^2=M_{1,0}^2+32\pi^2\zeta v^2/N_1$, and including it gives
\begin{equation}\label{eq:depth}
\Delta f_*=-0.0029673\,n^5T^4\Big[1-\frac{1.817}{\sqrt n}+O\Big(\frac{\log n}{n}\Big)\Big]\,,
\end{equation}
so the correction is suppressed by $n^{-1/2}$. All other corrections are suppressed by further powers of $1/n$, up to logarithms (End Matter).

Corrections could also change the shape of the condensate. A condensate of rank $N_2-1$, for which $\det\Phi_2=0$, differs in free energy from the full-rank one only by a relative amount of order $1/N_2$, which is smaller than the correction above. But the large sector couples to $\Phi_2$ only through $\tr\Phi_2^\dagger\Phi_2$, so integrating it out, to all orders in its own interactions, cannot distinguish condensates with the same trace. The single-trace term $\htl_2\tr(\Phi_2^\dagger\Phi_2)^2$ then gives every stationary point of rank $0<k<N_2$ a direction of negative curvature, $-2\htl_2v_k^2$, where $v_k^2$ is the nonzero eigenvalue of $\Phi_2^\dagger\Phi_2$ there (End Matter).

\emph{Entropic order.}---It remains to understand why heating does not restore the symmetry. In any homogeneous phase of a CFT the free energy density $f$ is proportional to $T^4$, leading to entropy and energy densities $s=-4f/T$ and $e=-3f$. The ordered state therefore has both more entropy and more energy than the symmetric state, $\Delta s_*=-4\Delta f_*/T$ and $\Delta e_*=-3\Delta f_*$: order at all temperatures in a CFT is necessarily entropic. Which modes supply this entropy can be read off directly from $c_2$. Each bosonic mode of mass $M$ contributes $T^2M^2/24$ to $f$, so $c_2$ records how the condensate shifts the masses of all modes: modes that become heavier contribute less entropy, and modes that become lighter add it. In Eq.~(\ref{eq:c2}), the terms $\frac12-\frac{\Rh}{6}\cos\theta$ come from the Higgsed vectors and the fluctuations of $\Phi_2$, which become heavier. The term $\frac{\Rh}{6}\sin\theta/r$, which is negative on the ordered arc, comes from the $2N_1^2$ real components of $\Phi_1$, whose mass squared decreases by $32\pi^2|\zeta|v^2/N_1$. The shift is suppressed by $1/N_1$, but it is shared by $N_1^2$ modes, so it governs the small-$r$ behavior, and the bound $r<0.19$ counts entropy. The disordered large sector serves as an entropy reservoir for the small sector, which orders. A single irreducible scalar multiplet cannot play both roles, since its thermal mass at one loop is nonnegative whenever its potential is bounded from below: gauge and Yukawa contributions are positive, and by irreducibility the scalar contribution is proportional to the average of the quartic potential over field directions. The mechanism is reminiscent of the Pomeranchuk effect in helium-3, where the solid has more entropy than the liquid~\cite{Pomeranchuk:1950}.

\emph{Superconductivity at parametrically high temperature.}---A superfluid becomes a superconductor when its broken symmetry is gauged. We can gauge $U(1)_{\Phi_2}$ by coupling the theory to a dynamical photon with a small gauge coupling $e$. The condensate Higgses $U(1)_{\Phi_2}$, and static magnetic fields are screened by the Meissner effect. In an ordinary Abelian plasma, there is electric Debye screening, but the static transverse photon has no perturbative screening mass. In the ordered phase, the condensate instead gives the photon a Meissner mass,
\begin{equation}\label{eq:meissner}
m_{\rm M}^2=2e^2N_2v^2\,,\qquad \frac{m_{\rm M}}{T}\simeq0.257\,e\,n^2\,,
\end{equation}
where we used $v^2/T^2\simeq0.0331\,n^3$ along the family. We keep $m_{\rm M}/T$ of order one and the gauge coupling small by defining $e=\widehat e/n^2$ at a reference scale $\mu_0$, with $\widehat e$ of order one. Then $m_{\rm M}/T\simeq0.257\,\widehat e$ stays finite, while the photon loop parameter $N_2^2e^2/16\pi^2$ vanishes as $n^{-2}$. The photon does not disturb the ordered fixed point, because its contributions to the thermal mass of $\Phi_2$ and the matter $\beta$-functions are subleading in $n$.

Unlike the matter sector, the weakly gauged theory is not conformal, since the gauge coupling runs according to
$\frac{de}{d\log\mu}=\frac{N_2^2e^3}{48\pi^2}$.
For $e(\mu_0)=\widehat e/n^2$, the corresponding Landau pole at $\Lambda_{\rm L}=\mu_0\exp\!\left(\frac{24\pi^2n^2}{\widehat e^2}\right)$ marks the breakdown of the weakly coupled description. Far below $\Lambda_{\rm L}$, the running is parametrically slow. Between $\mu_0$ and $\mu_0\exp(n)$, the gauge coupling changes only by $\frac{\Delta e}{e}=\frac{\widehat e^{\,2}}{48\pi^2n}+O(n^{-2})$.
The description remains controlled throughout this exponentially large range, and each gauged $\cT_n$ remains in the superconducting phase. Superconductivity from entropic order was previously proposed in a large-$N$ scalar model below its cutoff~\cite{Han:2026}; here the ordered sector is a CFT, and only the photon has a Landau pole.

\emph{Summary.}---The circle of planar fixed points of CCR is lifted at order $1/N^2$, and the ranks and numbers of flavors can be chosen such that a fixed point remains on the ordered arc. Since all couplings along the family $\cT_n$ are of order $1/n$, the existence of the fixed point and the superfluidity in its thermal state are under parametric perturbative control. To our knowledge, the theories $\cT_n$, together with the theory of Komargodski and Popov~\cite{NoteKP}, are the first unitary, local, four-dimensional CFTs with finitely many fields whose thermal state breaks a continuous global symmetry spontaneously at every temperature. In addition, coupled to a weakly coupled photon, the same theories are superconductors over an exponentially large range of temperatures. Ordered fixed points also exist in much smaller theories, such as the two examples above, but their thermal corrections are under far less control. The fixed points are multicritical: besides the two mass terms, three combinations of the quartic couplings are weakly relevant, so reaching them from a generic ultraviolet theory requires tuning.

\begin{acknowledgments}
This work made substantial use of the AI system Claude (Opus 5.5, Anthropic), accessed through the Claude Science research environment. We thank Z.~Komargodski and F.~K.~Popov for discussions and for sharing their unpublished results, M.~Smolkin and E.~Rabinovici for comments on the manuscript, and T.~Dumitrescu and D.~Delmastro for useful discussions.
\end{acknowledgments}

\input{refs_prl}
\section*{End Matter}
\appendix

\begin{table*}[t]
\caption{Fixed points of the one- and two-loop $\beta$-functions, exact in $N_i$. Here $\theta^{\rm 1L}$ and $\theta^{\rm 2L}$ are the angles $\theta=\arg[(f_1-f_2)/2+i\zeta]$ at one and two loops, $c_2^{\rm 2L}$ is evaluated with the two-loop couplings, and $D/D_{\rm LO}$ is the depth of the ordered minimum including the correction from the static modes of the large sector, relative to its leading-order value.}
\label{tab:fp}
\begin{ruledtabular}
\begin{tabular}{cccccccc}
theory & $10^3\lambda_1^*$ & $10^3\lambda_2^*$ & $\Delta$ & $\theta^{\rm 1L}$ & $\theta^{\rm 2L}$ & $c_2^{\rm 2L}/\lambda_2^*$ & $D/D_{\rm LO}$\\
\hline
\input{tab_fixedpoints_body}
\end{tabular}
\end{ruledtabular}
\end{table*}

\section{Finite-\texorpdfstring{$N$}{N} \texorpdfstring{$\beta$}{beta}-functions}

With $\epsilon_i=1/N_i^2$, $\epsilon_{12}=1/(N_1N_2)$, and $j\neq i$, the $\beta$-functions are as follows, exact in $N_i$. Through two loops the gauge couplings obey
\begin{equation}\label{eq:beta-gauge}
\beta_{\lambda_i}=-\frac{21-4x_i}{3}\,\lambda_i^2+\Big[\frac{26x_i-54}{3}-(4+2x_i)\,\epsilon_i\Big]\lambda_i^3\,,
\end{equation}
whose zero is Eq.~(\ref{eq:BZ}). The quartic couplings have $\beta=\beta^{\rm 1L}+\beta^{\rm 2L}$, with
\begin{equation}\label{eq:betas-1loop}
\begin{split}
\beta^{\rm 1L}_{h_i}={}&8h_i^2-12(1-\epsilon_i)\lambda_ih_i+\big(\tfrac32-12\epsilon_i\big)\lambda_i^2+24\epsilon_if_ih_i\,,\\
\beta^{\rm 1L}_{f_i}={}&4(1+4\epsilon_i)f_i^2+16f_ih_i-12(1-\epsilon_i)\lambda_if_i+12h_i^2\\
&+\big(\tfrac92+6\epsilon_i\big)\lambda_i^2+4\zeta^2\,,\\
\beta^{\rm 1L}_{\zeta}={}&\zeta\big[4(1+\epsilon_1)f_1+4(1+\epsilon_2)f_2+8h_1+8h_2\\
&-6(1-\epsilon_1)\lambda_1-6(1-\epsilon_2)\lambda_2+8\epsilon_{12}\zeta\big]\,,
\end{split}
\end{equation}
and
\begin{equation}\label{eq:betas-2loop}
\begin{split}
\beta^{\rm 2L}_{h_i}={}&\big(4-\tfrac83x_i+\tfrac23(15+32x_i)\epsilon_i-120\epsilon_i^2\big)\lambda_i^3 \\
&+\big(\tfrac{20}{3}x_i-26-\tfrac43(9+5x_i)\epsilon_i+18\epsilon_i^2\big)\lambda_i^2h_i \\
&+12\epsilon_i(1-8\epsilon_i)\lambda_i^2f_i+8(5-11\epsilon_i)\lambda_ih_i^2 \\
&+96\epsilon_i(1-\epsilon_i)\lambda_ih_if_i-24(1+5\epsilon_i)h_i^3 \\
&-352\epsilon_ih_i^2f_i-8\epsilon_i(5+41\epsilon_i)h_if_i^2-40\epsilon_ih_i\zeta^2\,,\\[2pt]
\beta^{\rm 2L}_{f_i}={}&\big(12-8x_i+\tfrac23(51-16x_i)\epsilon_i+60\epsilon_i^2\big)\lambda_i^3 \\
&+\big(\tfrac{20}{3}x_i-14+\tfrac{20}{3}(6-x_i)\epsilon_i+58\epsilon_i^2\big)\lambda_i^2f_i \\
&+4(9-4\epsilon_i)\lambda_i^2h_i+48(2-\epsilon_i)\lambda_ih_i^2 \\
&+128(1-\epsilon_i)\lambda_ih_if_i+16(2+3\epsilon_i-5\epsilon_i^2)\lambda_if_i^2 \\
&+32(1-\epsilon_j)\lambda_j\zeta^2-96h_i^3-8(5+41\epsilon_i)h_i^2f_i \\
&-352\epsilon_ih_if_i^2-24\epsilon_i(3+7\epsilon_i)f_i^3 \\
&-40\epsilon_if_i\zeta^2-32\epsilon_{12}\zeta^3\,,\\[2pt]
\beta^{\rm 2L}_{\zeta}={}&\zeta\sum_i\Big[\big(\tfrac{10}{3}x_i-7+\tfrac23(3-5x_i)\epsilon_i+5\epsilon_i^2\big)\lambda_i^2 \\
&\quad+64(1-\epsilon_i)\lambda_ih_i+32(1-\epsilon_i^2)\lambda_if_i \\
&\quad-20(1+\epsilon_i)h_i^2-80\epsilon_ih_if_i-20\epsilon_i(1+\epsilon_i)f_i^2\Big] \\
&+\epsilon_{12}\zeta^2\sum_i\big[8(1-\epsilon_i)\lambda_i-96h_i-48(1+\epsilon_i)f_i\big] \\
&-4(\epsilon_1+\epsilon_2+8\epsilon_1\epsilon_2)\zeta^3\,.
\end{split}
\end{equation}
Both contain only the corrections $\epsilon_i$ and $\epsilon_{12}$, and no positive powers of $N_1/N_2$. The two-loop terms follow from the general two-loop results~\cite{MachacekVaughn:1985,LuoWangXiao:2003}, and we checked all of these $\beta$-functions against PyR@TE~3~\cite{SartoreSchienbein:2021} for several values of the ranks. Since $\beta_\zeta\propto\zeta$, the decoupled points $\zeta=0$ are exact fixed points.

\section{The \texorpdfstring{$\beta$}{beta}-function on the circle}

Write the quartic couplings as $q=(h_1,h_2,f_1,f_2,\zeta)$ and split $\beta=\beta^{(0)}+\beta^{(1)}$, where $\beta^{(0)}$ is the planar part at $\lambda_1=\lambda_2=\lb$, and $\beta^{(1)}$ contains the terms with $\epsilon_i$ and $\epsilon_{12}$ and the terms due to $\delta$. Near a point $q_0(\theta)$ of the circle, a shift $q=q_0+\delta q$ changes $\beta^{(0)}$ by $J\,\delta q$, where $J$ is its Jacobian. Exactly one combination of the $\beta$-functions is unchanged at this order, $\ell\cdot J=0$, with
\begin{equation}
\ell(\theta)=\big(-p\sin\theta,\,p\sin\theta,\,-\sin\theta,\,\sin\theta,\,2\cos\theta\big)\,,
\end{equation}
and $p=(2+3\sqrt6)/4$. All other combinations, the transverse $\beta$-functions, can be set to zero by a choice of $\delta q$, and the fixed-point condition reduces to $\ell\cdot\beta^{(1)}(q_0(\theta))=0$. Since $\ell\cdot\partial_\theta q_0=2\Rh\lb$, the $\theta$-component of $\beta^{(1)}$ is $\beta^{(1)}_\theta=\ell\cdot\beta^{(1)}(q_0(\theta))/(2\Rh\lb)$, the leading term of $\beta_\theta=d\theta/d\log\mu$. Every simple zero of $\beta^{(1)}_\theta$ gives a fixed point near $q_0(\theta)$, by the implicit function theorem. Inserting~(\ref{eq:betas-1loop}) we find
\begin{equation}
\begin{split}
\beta^{(1)}_\theta=\frac{\sin\theta}{2\Rh}\,\Big\{&\kappa\,\delta+\lb\Big[(\epsilon_2-\epsilon_1)\big(\kappa\cos^2\theta+a\big)\\
&+(\epsilon_1+\epsilon_2)\,b\cos\theta+\epsilon_{12}\,\kappa\sin2\theta\Big]\Big\}\,,
\end{split}
\end{equation}
with $\kappa=(18\sqrt6-39)/2=8\Rh^2$, $a=\frac{93}{2}-\frac{27\sqrt6}{4}\approx29.966$, and $b=\sqrt{6\sqrt6-13}\,\big(\frac{9\sqrt3}{2}-\frac{135\sqrt2}{8}\big)\approx-20.935$. This is Eq.~(\ref{eq:L}) with
\begin{equation}\label{eq:gr}
g_r(\theta)=(1-r^2)\big(\kappa\cos^2\theta+a\big)+(1+r^2)\,b\cos\theta+r\kappa\sin2\theta\,.
\end{equation}
Every term is of order $1/N^2$: the non-planar terms through $\epsilon_i$ and $\epsilon_{12}$, and the mismatch through $\Delta$. At the interacting zeros the slope $\partial_\theta\beta^{(1)}_\theta=\lb\sin^2\theta_*|2\kappa\cos\theta_*+b|/(2\Rh N_2^2)>0$ gives the dimension $4+\partial_\theta\beta^{(1)}_\theta$ of the operator that moves the fixed point along the old circle.

\section{Fixed points}

Table~\ref{tab:fp} lists the fixed points of the one- and two-loop $\beta$-functions, exact in $N_i$, for the two examples and for members of the family~(\ref{eq:family}). In all cases the potential is bounded below, $c_1/\lambda_1^*$ lies between $0.48$ and $0.56$, $c_2<0$, and the leading-order minimum~(\ref{eq:min}) is a condensate of full rank.

\section{Thermal corrections}

The leading correction to the thermal free energy comes from the static modes of the $2N_1^2$ real scalars of the large sector, whose mass in the background $\Phi_2=v\mathbf 1$ is $M_1^2=M_{1,0}^2+32\pi^2\zeta v^2/N_1$. With $w=v^2/(N_1T^2)$ and the limiting ratios $\widehat a=a_2/\lb$, $\widehat\zeta=\zeta/\lb$, and $\widehat c=c_1/\lb$, the free energy density along the full-rank direction is
\begin{equation}
\begin{split}
\frac{\Delta f(w)}{16\pi^2\lb N_1^2T^4}={}&\widehat a\,w^2+\frac{\widehat\zeta}{6}\,w\\
&+\frac23\sqrt{\lb}\,\Big[\widehat c^{\,3/2}-\big(\widehat c+2\widehat\zeta w\big)^{3/2}\Big]+\cdots\,.
\end{split}
\end{equation}
At leading order the minimum is at $w_0=-\widehat\zeta/(12\widehat a)\approx0.033116$, with depth $D_{\rm LO}=\widehat a\,w_0^2\approx1.5502\times10^{-3}$. The last term reduces the depth to $D/D_{\rm LO}=1-16.503\sqrt{\lb}+O(\lb)$, which with $\lb\simeq2/(165n)$ gives Eq.~(\ref{eq:depth}). The last column of Table~\ref{tab:fp} uses the exact two-loop couplings and includes the contributions of the small sector to $c_2$. The remaining corrections can be estimated by power counting. Loops of the small sector carry no factor of $N_1/N_2$ and are suppressed by $r=1/n^2$; higher loops of the large sector are suppressed by $\lb$, up to logarithms of $M_W/T\sim\sqrt n$; and the soft modes of the unbroken gauge groups enter only through further loops. All are therefore of relative order $\log n/n$ or smaller.

For the shape of the condensate, let $G(S)$ collect all terms of the effective potential that depend on $\Phi_2$ only through $S=\tr\Phi_2^\dagger\Phi_2$: the double-trace term, the thermal mass, and the contribution of the large sector to all orders in its interactions, so that $V_{\rm eff}=\htl_2\tr(\Phi_2^\dagger\Phi_2)^2+G(S)+(\text{small-sector loops})$. At a stationary point of rank $k$, with $k$ equal nonzero eigenvalues $v_k^2$, stationarity gives $G'(S)=-2\htl_2v_k^2$, which is the curvature in the direction of a vanishing eigenvalue, so every such point is a saddle, independently of $G$. The small-sector loops that could distinguish shapes are suppressed by $\lb$, up to logarithms, and carry no power of $N_1/N_2$.

\end{document}

%% file: tab_fixedpoints_body.tex
$(150,783,5,26)$ & 1.468 & 2.517 & $-33.47$ & $-1.844$ & $-1.700$ & $-0.793$ & 0.221\\
$(300,1562,10,52)$ & 2.130 & 2.476 & $-38.25$ & $-1.957$ & $-1.828$ & $-1.79$ & 0.261\\
\hline
$\mathcal{T}_{5}$ & 1.271 & 2.517 & $-41.85$ & $-2.230$ & $-2.143$ & $-0.588$ & 0.200\\
$\mathcal{T}_{9}$ & 1.156 & 1.368 & $-34.60$ & $-1.796$ & $-1.728$ & $-5.76$ & 0.477\\
$\mathcal{T}_{13}$ & 0.871 & 0.941 & $-33.20$ & $-1.725$ & $-1.678$ & $-13.9$ & 0.555\\
$\mathcal{T}_{21}$ & 0.564 & 0.580 & $-32.43$ & $-1.687$ & $-1.659$ & $-39.2$ & 0.641\\
$\mathcal{T}_{41}$ & 0.294 & 0.296 & $-32.06$ & $-1.670$ & $-1.655$ & $-155$ & 0.735\\
$\mathcal{T}_{101}$ & 0.120 & 0.120 & $-31.93$ & $-1.664$ & $-1.658$ & $-951$ & 0.827\\